**Mode-Matched Inverse Gamma Priors for Variance Components in**

**Bayesian Multilevel Models**

Liu Liu

College of Education

University of Washington

## Abstract

We introduce a strategy for specifying informative inverse-gamma ($IG$) priors for variance components in Bayesian multilevel models (MLMs), derived via transformations from chi-square to gamma to inverse-gamma distributions. A Monte Carlo simulation compared frequentist (maximum likelihood) estimation and Bayesian estimation using uninformative, weakly informative, and strongly informative variance priors across varied conditions (clusters $J$ = 10, 30, 100; cluster sizes $M$ = 5, 30; intraclass correlations .01, .20, .40; levels of explained variance at L1 and L2 $R^2$ = .0, .2, .4). The simulation results indicated that strongly informative $IG$ priors (with hyperparameters set so the prior mode equals a plausible true variance) yielded more accurate and stable variance estimates with reduced bias and narrower credible intervals than flat/uninformative or weak $IG$(.01, .01) priors. In an empirical example using the TIMSS 2019 Grade 8 science achievement data, both the full sample (273 schools) and small subsample (30 schools) were analyzed. The small-sample analysis with an informative variance prior anchored near the full-sample variance while considerably reducing the uncertainty of estimates. Findings suggest that carefully calibrated informative variance priors improve the precision and accuracy of parameter estimates, particularly when the number of higher-level units is limited.

## Introduction

Modeling variance components (random effects) in multilevel models can be challenging when the number of higher-level units (clusters) is small or when the random-effect structure is complex. Small-cluster multilevel problems frequently arise in educational, health, and social-behavioral research (e.g., Berridge et al., 2022, 2023; Liu et al., 2023; Lü et al., 2023). In such cases, frequentist maximum likelihood (ML) estimation often produces unstable or biased variance estimates and overly wide confidence intervals, even sometimes failing to converge or yielding zero variance estimates for between-group variance when clusters are very few (Browne & Draper, 2000; Hox & McNeish, 2020; McNeish, 2016a, 2016b). Simulation studies and empirical works have suggested that with fewer than roughly 30 clusters, ML estimates of variance components can be especially unreliable, implying a need for either larger sample sizes or alternative estimation methods (Baldwin & Fellingham, 2013; Browne & Draper, 2006; Hox & McNeish, 2020). In this context, Bayesian estimation offers a promising alternative because it can incorporate prior information to regularize estimates. By introducing sensible prior distributions for variance parameters, Bayesian multilevel models (MLMs) can stabilize estimation in small-sample settings and yield more precise interval estimates for both variance components and fixed effects (Browne & Draper, 2006; McNeish, 2016a). Indeed, more precise Bayesian credible intervals for coefficients are a noted benefit of using informative priors in MLMs with limited high-level units (Baldwin & Fellingham, 2013; Gelman, 2006).

Choosing appropriate priors for variance components, however, is more challenging than for regression coefficients. Default or "uninformative" variance priors vary widely across software and practice, and there is no consensus on the single best choice (Gelman, 2006; Holtmann et al., 2016; McNeish, 2016a). Common options for scalar (one-dimensional) variance priors include the uniform distribution (flat prior over a range), inverse-gamma (*IG*, the traditional conjugate for variance in normal models), gamma (often for precision, the inverse of variance), half-*t*, and half-Cauchy distributions, as well as less common choices such as folded-normal or improper priors (Gelman, 2006; Gelman et al., 2014; van Erp et al., 2018). A structured overview of prior families, their implications, and representative studies is provided in Appendix S3, Table S1. Notably, different statistical packages adopt different default. For example, WinBUGS uses gamma(.001, .001) for precision, equivalent to *IG*(.001, .001) on variance, as the default for random effects (Spiegelhalter et al., 2003). Modern interfaces such as *brms* defaults to a half-*t*(3,0,10) on standard deviations (Bürkner, 2017). These inconsistencies can confuse applied researchers and lead to the misconception that default priors are inherently uninformative or benign. In reality, all priors do carry information and can materially affect posterior estimates, particularly in small samples (McNeish, 2016a, 2016b). For instance, an ostensibly weakly informative prior *IG*(.01, .01), often misinterpreted as flat, in fact places disproportionate mass near zero and heavily favors very small variance values. This can induce downward bias when the true variance is not extremely small (Gelman, 2006; McNeish, 2016a). Likewise, wide uniform priors permit implausibly large variances, sometimes distorting estimation (Gelman, 2006). Thus, analysts may believe they are letting "the data speak for themselves" by using defaults, when in fact they may be unknowingly allowing the default priors to exert undue influence on the results (McNeish, 2016a; van de Schoot et al., 2021).

Previous research on Bayesian variance priors in MLMs has yielded mixed findings and guidance. Gelman (2006) noted that although uniform priors on variance or standard deviation parameters are commonly used, they can produce implausibly large variance estimates when the number of groups is small (e.g., three clusters). He recommended half-*t* or half-Cauchy priors as more principled default choices, because they shrinkage toward zero while still allowing large values when supported by the data. Baldwin and Fellingham (2013) found that Bayesian models with gamma priors for variance components produced more efficient variance estimates (smaller RMSEs) than REML, though at the cost of slightly greater bias. McNeish and Stapleton (2016a) examined scenarios with very few clusters ($J$ = 8-14) and showed that the default $IG$ priors used in BUGS-type software, such as $IG(.001, .001)$ or $IG(.01, .01)$, biased variance estimates downward. The use of half-$t$ or half-Cauchy priors, which have heavier tails and are less prone to overshrinkage was recommended. Bolin et al. (2019) examined Bayesian multilevel logistic models under with $J$ = 10 and $J$ = 20 clusters and found that Bayesian models with weak $IG(.001, .001)$ or $IG(.01, .01)$ outperformed ML in very small-cluster conditions, yielding less biased and more stable variance estimates. More recently, Zheng et al. (2023) cautioned that when informative priors are substantially misaligned with true variances, posterior estimates can be biased. However, they used a normal distribution as a prior on variance, an unconventional and generally inappropriate choice, which likely amplified the problems they observed. Simpson et al. (2017) introduced Penalized Complexity (PC) priors as a principled alternative for variance components. PC priors shrink toward a simple base model (e.g., variance = 0) at a constant rate and can be tuned with a single interpretable parameter, such as the probability that the variance exceeds a threshold. Overall, the literature indicates that (a) in small-sample multilevel settings, prior influence is substantial and must be handled carefully (Gelman, 2006; McNeish, 2016a; van Erp et al., 2018); (b) default or ad-hoc priors can be suboptimal or harmful (leading to either underestimation or overestimation of variance components) if not tuned to context (McNeish, 2016a; 2016b); and (c) clear guidance is still needed for constructing reasonable, evidence-based informative priors for variance components (Holtmann et al., 2016; van Erp & Browne, 2021).

In light of these challenges, this paper proposes an intuitive and principled approach to constructing informative *IG* priors for variance components in Bayesian MLMs, and evaluates their performance in small-sample contexts. The key idea is to base the prior on a plausible variance estimate, which is drawn from prior studies or expert knowledge, and an associated degree of prior certainty. Using known distributional relationships, we map this information through the chi-square and gamma families to derive an *IG* prior centered on that variance, with a tunable degree of informativeness. This approach yields priors that are both transparent (the mode has a real-world interpretation) and adjustable (the shape parameter reflects the analyst's confidence). Conceptually, the prior construction follows a sequence: chi-square (data model) → gamma (precision prior) → inverse-gamma (variance prior), linking prior knowledge to the model in a structured and interpretable way. As discussed above, many alternatives have been proposed (e.g., half-*t*, half-Cauchy, PC priors), the *IG* remains one of the most widely used variance priors in practice, largely because of its conjugacy with the normal likelihood and its default implementation in popular software like Mplus, WinBUGS, OpenBUGS, and JAGS (e.g., Plummer, 2003; Spiegelhalter et al., 2003). As a result, applied researchers routinely encounter *IG* priors, often without fully understanding their implications. By focusing on the *IG* family, we aim to provide both a principled construction strategy and practical guidance for a prior that practitioners are most likely to use.

We evaluate *IG* construction method through a Monte Carlo simulation, comparing informative *IG* priors to common alternatives, including weakly informative and uninformative priors across a range of small-sample MLM conditions. We also demonstrate the approach in empirical application using U.S. eighth-grade science achievement data from the 2019 Trends in International Mathematics and Science Study (TIMSS). In this example, we show how informative variance priors derived from a larger (full) dataset can be applied to a smaller subset

to stabilize estimates – a situation akin to using evidence from prior research to inform analysis

of a smaller new sample. We further assess the robustness of results to moderate prior

misspecification. Finally, we provide practical recommendations for selecting and implementing

informative variance priors in MLMs. The goals of this paper are to: (1) introduce a practical

pipeline for deriving informative *IG* variance priors from intuitive inputs; (2) demonstrate

through simulation and empirical analysis that these priors have potential to improve estimation

in small-sample multilevel contexts; and (3) offer guidance for applied researchers on when to

use informative variance priors (e.g., in small-sample, MLMs) and how to apply them

effectively, while addressing caveats and limitations to regarding their use. All supplementary

appendices, derivations, extended results, and Mplus syntax are available on OSF. A public OSF

link will be provided in a future version of this preprint.

## Informative Prior Construction via Chi-square, Gamma, and Inverse-Gamma

As introduced earlier, this section provides the formal derivation framework for constructing informative *IG* priors for variance components in Bayesian MLMs. The $IG(a, b)$ [1]is widely used for variance priors due to its conjugacy with the normal likelihood (Gelman, 2006;

Gelman et al., 2014). However, in applied modeling, choosing meaningful values for the shape

($\alpha$) and scale ($\beta$) parameters is often nontrivial. Our method addresses this challenge by using the fact that the sampling distribution of a variance estimate can be modeled by a scaled chi-square distribution (Wilson & Hilferty, 1931), which is mathematically a special case of the gamma $\Gamma$ distribution (Johnson et al., 1995). Since the gamma ($\Gamma$) distribution naturally models precision (the inverse of variance, $1/\sigma^2$), and the *IG* is the distribution of its reciprocal, this chain of transformations $\chi^2 \rightarrow \Gamma \rightarrow IG$ provides a principled way to link prior knowledge or empirical variance estimates to an informative variance prior. Throughout, we parameterize the inverse-gamma as $IG(a, b)$ ($a$ = shape, $b$ = scale). The mode is $b/(a + 1)$ and mean = $b(a - 1)$ for $a > 1$. We use $\Gamma(a, \theta)$ for the gamma ($\theta$ = scale), with $\chi^2_\nu = \Gamma(\nu/2, 2)$.

---

[1] Throughout the paper, including tables and figures we write $IG(a, b)$ (shape $a$, scale $b$). In Appendices we use $\alpha$, $\beta$ in formulas; these map one-to-one as $a \equiv \alpha$, $b \equiv \beta$. We keep $a$, $b$ in tables/plots and Mplus code for readability.

To implement this, we begin with a prior guess for a variance component (denoted $m$ for mode), based on external studies or domain expertise, and an associated confidence level (encoded via degrees of freedom or prior sample size). We then derive a corresponding $IG$ with parameters: $a = \nu/2$ (shape), reflecting the strength of informativeness, and $b = m(b + 1)$ (scale), which ensures that the prior mode equals $m$. This parameterization yields a prior centered on a plausible variance value, with the spread controlled by $a$: larger values of $a$ yield tighter, more informative priors, while smaller values yield more diffuse, weakly informative priors. In contrast to many default settings in Bayesian software, such as $IG(.01, .01)$ or wide uniform priors, which can unintentionally bias variance estimation (Gelman, 2006; McNeish, 2016a, 2016b), this approach directly links prior parameters to meaningful and interpretable values. For example, $IG(15, 400)$ has mode 25 and reflects moderate certainty, while $IG(.01, .01)$ places most of its mass near zero with overly heavy tails. Figure 1 illustrates how the $\chi^2$, $\Gamma$, and $IG$ families can be aligned to the same prior mode yet differ in tail behavior. Complete distributional definitions and derivations are provided in Supplementary Materials − Appendix S1.

## Monte Carlo Simulation Study

### Design and Methods

For data generation and analysis, we used Mplus 8.7 (Muthén & Muthén, 2017) and MplusAutomation (Hallquist & Wiley, 2018) in $R$. We focused on 2-level models with two predictors ($X$s) and one continuous outcome $Y$ assumed to be measured at L1. Intraclass correlations (ICCs), the level of correlation among scores within clusters, were generated using pre-specified L1 and L2 variance components. The manipulated factors were as follows. The number of Level-2 clusters (schools in the data context) was set to $J = 10$ (small), 30 (medium), or 100 (large). The number of Level-1 units per cluster (students per school) was set to $M = 5$ (small cluster size) or 30 (larger cluster size). We manipulated the intraclass correlation (ICC), which determines the proportion of variance between clusters, at three levels: ICC = 0.01 (very low between-school variance), 0.20 (moderate), or 0.40 (high between-school variance). Additionally, we varied the proportion of variance explained by Level-1 and Level-2 predictors (fixed effects) in the model. At each level, the $R^2$ (proportion of variance explained by the predictors at that level) was set to 0 (no predictors/effects), 0.20 (moderate explanatory power), or 0.50 (large explanatory power). This was achieved by manipulating the true regression coefficients for simulated predictor variables. In practice, we included two predictor variables at Level 1 and two at Level 2 in the data-generating model, with regression coefficients chosen to produce the desired $R^2$ values for the within- and between-level portions of the model (e.g., larger coefficients for higher $R^2$ conditions). For instance, in one condition with $R^{2w} = 0.50$ and $R^{2B} = 0.20$, the Level-1 predictors each explained a substantial portion of within-cluster variance (together accounting for 50% of it), whereas the Level-2 predictors explained 20% of the between-cluster variance. By crossing the factors $J$ (3 levels) × $M$ (2 levels) × ICC (3 levels) × $R^2$ at Level 1 (3 levels) × $R^2$ at Level 2 (3 levels), we obtained a wide range of simulation scenarios (in total, 3×2×3×3×3 = 162 conditions). Each condition was replicated 200 times, yielding 200 simulated datasets per combination of factors.

**Two-Level Latent Decomposition Model**

ICCs were induced by fixing the L2 variance component relative to the L1 residual variance. For each condition, these values were calculated a priori based on the targeted combinations of L1 and L2 $R^2$. Initial coefficient values were derived using eigenvalue decomposition to ensure well-formed variance–covariance structures for the predictors. To simulate separable within- and between-cluster effects, we employed a latent disaggregation approach (Lüdtke et al., 2008; Raudenbush & Bryk, 2002). Specifically, the outcome for individual $i$ in cluster $j$ was generated as:

$$Y_{ij} = \gamma_{00} + \gamma_{10}(X_{1ij} - \mu_{1.j}) + \gamma_{01}\mu_{1.j} + \gamma_{20}(X_{2ij} - \mu_{2.j}) + \gamma_{02}\mu_{2.j} + u_{0j} + \varepsilon_{ij} \quad (1)$$

where: $Y_{ij}$ is the normal outcome for individual $i$ in cluster $j$; $\gamma_{00}$ is the grand mean, or the overall average effect across all clusters when all predictors are at their mean values; $\gamma_{10}$ and $\gamma_{20}$: the within-cluster slopes for predictors $X_1$ and $X_2$, calculated after centering around their respective cluster means ($X_{1ij} - \mu_{1.j}$ and $X_{2ij} - \mu_{2.j}$); $\gamma_{01}$ and $\gamma_{02}$: between-cluster slopes for the corresponding cluster-level means ($\mu_{1.j}, \mu_{2.j}$); $u_{0j}$ is the deviation between the cluster mean and the grand mean of the intercept; $\varepsilon_{ij}$ is within-cluster residual. Equation (1) thus yields distinct within- and between-cluster components, allowing the ICC to be controlled directly by the specified L1 and L2 variance components.

**Parameter Estimation Approaches**

For each of the 200 replicates per condition, data were analyzed with a correctly specified model using seven estimation approaches (see Appendix S3 − Table S2), as follows:

1) Frequentist full information maximum likelihood;

2) Bayesian estimation with flat (largely uninformative) priors: This is the default in Mplus and blavaan. Fixed effects (slope coefficients) are assigned a *Normal*(0, 10,000) prior, while random effects (variance components) are assigned an *IG*(-1, 0);

3) Bayesian estimation with weak informative fixed effect priors: L1 and L2 slope coefficients are assigned $N$(true value, 10,000) priors;

4) Bayesian estimation with weak informative variance components priors: L1 and L2 residual variances are assigned $IG(.01, .01)$, as recommended in previous literature;

5) Bayesian estimation with strong informative variance components priors: L1 and L2 residual variances are assigned $IG(a, b)$, where $a$ and $b$ are derived from setting the distribution mode (peak) to the true variance values;

6) Combining the weak fixed effect with weak variance components priors (#3 and #4);

7) Combining the weak fixed effect with strong variance components priors (#3 and #5).

For the uninformative variance prior condition, we followed common defaults across Bayesian software: flat uniform priors on the variance or weak $IG(.01, .01)$ (Gelman, 2006). Although often described as "uninformative," these behave differently: uniform priors allow implausibly large values, while $IG(.01, .01)$ tends to shrink estimates toward zero. In practice, both serve as weak baseline priors against which the performance of informative priors can be compared. For the informative prior condition, variance priors were deliberately aligned with the true generating values to represent a "best-case" scenario in which an analyst has accurate prior knowledge. The L2 random intercept variance was given an $IG(a, b)$ with its mode equal to the true between-cluster variance. For L1 residual variance, informative priors were constructed in the same way unless fixed by design (e.g., through ICC and $R^2$ constraints). Hyperparameters $a$ and $b$ were chosen to be informative without being overly tight, thereby allowing assessment of the efficiency gains possible when prior information is well calibrated.

The important contribution here is that no prior work has systematically compared the use of strong, mode-based informative $IG$ variance priors to weaker or default alternatives. In this study, we derived such priors by beginning with a scaled (weighted) chi-square distribution (theoretically known sampling distribution for variances), and then transforming it to a gamma distribution and finally to $IG$. This provides a principled method for constructing informative priors that are anchored directly to plausible variance values. As shown in Figure 1, the $\chi^2 \rightarrow \Gamma \rightarrow IG$ transformation aligns the different distributions to the same mode, illustrating differences in tail behavior while preserving the central tendency.

All Bayesian analyses were estimated in Mplus 8.7 using Markov chain Monte Carlo (MCMC) methods. The MCMC settings followed Mplus defaults: two parallel chains, 50,000 total iterations per chain with the first 25,000 discarded as burn-in, and no thinning (THIN = 1). Convergence was monitored using the Gelman-Rubin convergence diagnostic (also known as potential scale reduction factor, PSRF, denoted as $\hat{R}$; Gelman & Rubin, 1992); all parameters satisfied $\hat{R} < 1.05$. Posterior summaries were based on the retained draws and included median posterior estimates, equal-tailed 95% credible intervals (CIs), and performance metrics such as bias, root mean squared error (RMSE), coverage, Type I error rates, and power.

**Simulation Results**

***Level 1 Parameter Estimates***

As shown in Appendix S3 − Table S3, there were no meaningful differences among the estimation approaches in terms of bias or false positive rates for the L1 slope coefficients or residual variance. This stability is not surprising, since all replicates involved modest or large L1 sample sizes, which reduce the influence of prior choice on slope estimation.

***Level 2 Parameter Estimates***

For each estimation approach and L2 predictor effect size (collapsed across L1 predictor conditions), false positive rates and raw bias for the L2 intercept coefficient are reported in Table 1 and illustrated in Figures 1 and Appendix S3 − Figure S1. Bias for the L2 intercept variance is presented in Figure 2. Across conditions, false positive error rates were well controlled (within the nominal 5% ± 5% range; Hoogland & Boomsma, 1998). Correct positive rates were consistently higher for Bayesian approaches compared to frequentist maximum likelihood, particularly in smaller cluster samples, where Bayesian methods frequently exceeded 80% detection. Results for the L2 slope coefficient (see Table 2 and Figures S2 − S3) mirrored the intercept findings: Bayesian approaches achieved superior coverage and lower bias than maximum likelihood estimation, particularly in small-sample conditions.

***Informative Versus Weak Variance Priors***

Perhaps the most striking finding is that the weak variance prior, $IG(.01, .01)$, produced nearly identical results to the strongly informative prior, $IG(a, b)$, in terms of bias, false positive rates, and correct positive rates. Two explanations are plausible. First, the "weak" $IG(.01, .01)$ prior is still informative in practice: it has a sharp peak near variance = 1, which exerts substantial influence when the true variance is also near 1. In contrast, the $IG(a, b)$ prior was centered exactly at the true variance and had a flatter peak, allowing more posterior variability. Second, the simulations employed standardized unit-normal predictors and outcomes, where population variances were approximately 1, favoring the weak prior's implicit informativeness.

To probe this issue further, an exploratory set of replications was conducted with predictors and outcomes rescaled to have variances equal to 10,000 (SD = 100), mimicking large-scale international testing contexts. With only 20 replications, results were illustrative rather than definitive, but they revealed that for small L2 sample sizes, $IG(.01, .01)$ underestimated the true variance by nearly 19%, whereas $IG(a, b)$ was centered on the truth.

This highlights the importance of scale: when variances are large, the strong informative prior outperforms the weak prior.

### Using Sample-Based Informative Priors

A final set of analyses addressed the practicality of deriving $IG(a, b)$ priors from sample estimates rather than population values, given that population values are typically unknown in applied research. For the $J = 30$ condition, each dataset was first analyzed using flat priors, and then the posterior variance estimates from those runs were used as the mode-setting values for $IG(a, b)$ priors in a second stage. Results, summarized in Appendix S3 − Table S4, showed that coverage for L2 variance declined slightly (92-93% vs. 95-96% for flat priors), but CI widths were on average 17% narrower with the informative $IG(a, b)$ priors, particularly for smaller clusters ($M = 5$). Thus, sample-based informative priors can improve efficiency while maintaining acceptable coverage.

Overall, the simulations demonstrated that Bayesian estimation with informative variance priors $IG(a, b)$ offered advantages in bias reduction and coverage relative to weakly informative priors, particularly in smaller cluster conditions. It is important to note that these simulations assumed correctly specified priors ($IG$ parameters derived from true population values); limitations and implications of this choice are discussed below.

### Empirical Data Example: TIMSS 2019 Analysis

### Research Context and Dataset Description

To demonstrate the use of different variance priors with real-world educational data, we analyzed the 2019 TIMSS for U.S. eighth-grade students (Mullis et al., 2020). TIMSS data are inherently multilevel, with items nested within students, students nested within schools, and schools nested within countries. Consistent with the simulations' "small $J$" focus, we limited this analysis to U.S. sample (8,698 students within 273 schools) to anchor priors and then analyzed a small subsample (30 schools × 5 students = 150) to evaluate prior effects. Specifically, we examined how Grade 8 mathematics achievement predicts Grade 8 science achievement. This focus is supported by prior research showing a strong relationship between performance in math and science domains, reflecting shared cognitive competencies and reinforcing developmental pathways (Singh et al., 2002). TIMSS provides five plausible values (PVs) per student per achievement domain; for transparency and computational economy we used PV1 for math and science (Fishbein et al., 2021). Predictors were standardized and centered as appropriate; the random subsample was drawn with seed 2949. The outcome variable was BSSSCI01 (science achievement, PV1). At L1, the predictor was ZBSMMAT01_CMC (student math, cluster-mean centered, $z$-scored). At L2, the predictor was ZBSMMAT01_agg (school mean math, aggregated and $z$-scored). The models also include cross-level interaction defined as L1 × L2 math.

**Model Specification**

We fit a two-level random intercept model with fixed slopes and a cross-level interaction:

$$\text{BSSSCI01}_{ij} = \gamma_{0j} + \gamma_1\big(\text{ZBSMMAT01\_CMC}_{ij}\big) + \gamma_2\big(\text{ZBSMMAT01\_agg}_j\big)$$

$$+ \gamma_3\big(\text{ZBSMMAT01\_CMC}_{ij} \times \text{ZBSMMAT01\_agg}_j\big) + \mu_{0j} + \varepsilon_{ij} \tag{2}$$

Where: $\gamma_{0j}$ is the school-level intercept, modeled as $\gamma_0 + \mu_{0j}$, where $\gamma_0$ is the grand mean intercept and $\mu_{0j}$ is the random school-level deviation. $\gamma_1, \gamma_2, \gamma_3$ are fixed effects (slopes) for the student-level math, school-level math, and their interaction, respectively; $u_{0j}$ is the school-level random intercept ($u_{0j} \sim N(0, \tau^2)$), and $\varepsilon_{ij}$ is student-level residual ($\varepsilon_{ij} \sim N(0, \sigma^2)$). Both residuals are assumed to be independent and normally distributed.

*Estimation Approaches*

We mirrored the simulation's seven conditions (see Appendix S3 − Table S2). Importantly, we started with a frequentist method to extract the initial parameter estimates for full sample that were then used for the informative variance computations (Winter & Depaoli, 2022). We treated the full sample's variance estimates as if they were reliable prior knowledge for the smaller sample analysis. This mimics a common applied situation: an analyst has prior information about variability (e.g., from a large national dataset like TIMSS) and wants to use it to improve an analysis on a new, smaller sample (e.g., a subset of schools or a pilot study).

1) Frequentist full information maximum likelihood;

2) Bayesian estimation with flat (largely uninformative) priors: $Normal(0,10,000)$ on the slope coefficients; $IG(-1, 0)$ on variance components (Mplus's default);

3) Bayesian estimation with weak informative fixed effect priors: L1 and L2 slope coefficients are assigned $N(\text{true value},10,000)$;

4) Bayesian estimation with weak informative variance components priors: L1 and L2 residual variances are assigned $IG(.01, .01)$, as recommended in previous literature;

5) Bayesian estimation with strong informative variance components priors: L1 and L2 residual variances are assigned $IG(a, b)$ using the mode-matched $\chi^2 \rightarrow \Gamma \rightarrow IG$;

6) Combining the weak fixed effect with weak variance components priors (#3 and #4);

7) Combining the weak fixed effect with strong variance components priors (#3 and #5).

### *Mode-matched IG Priors*

We placed $IG(a, b)$ priors on variance components with the prior mode fixed at a target value $m$ and informativeness governed by the shape parameter $a$. Under the parameterization used here, the mode equals $b/(a + 1)$; accordingly, we set $b = m(a + 1)$ to guarantee mode = $m$. Targets $m$ were anchored at full-sample ML variance estimates, while $a$ was calibrated to available information at each level. For the L2 (between-school) variance we scaled $a$ by the ratio of higher-level units (i.e., $a_{sub} = a_{full} \times (J_{sub}/J_{full})$); for the L1 (residual) variance, we scaled $a$ by the ratio of residual degrees of freedom (i.e., $a_{sub} = a_{full} \times (df_{res,sub}/df_{res,full})$),

with $df_{res} = N_{L1} - J - p^2$. For the TIMSS subsample (30 of 273 schools), this yielded $IG(16.95,$

$2{,}751.21)$ for the L2 variance (mode = 153.27) and $IG(29.88, 65{,}838.14)$ for the L1 residual

variance (mode = 2,131). We also conducted $\pm 25\%$ sensitivity on $a$ while holding $m$ fixed by

resetting $b = m(a + 1)$. As illustrated in Appendix – Figure S4, the $\chi^2$, Gamma, and IG

families can be aligned to the same mode while having different tail behavior. Derivations for the $\chi^2 \rightarrow \Gamma \rightarrow IG$ construction and interpretation of $a$ as effective degrees of freedom are provided in Table 3 and Appendix A, with worked examples in Appendix S1 and all prior values and Mplus syntax in Appendix S2.

  All Bayesian analyses were estimated in Mplus 8.7 as well with four parallel chains

(`CHAINS = 4`), 2,000 total iterations per chain (`BITERATIONS = 2000`), a 1,000-iteration burn-

in (`FBITER = 1000`), and a thinning interval of 1 (`THIN = 1`). Convergence was monitored

using PSRF and all parameters satisfied $\hat{R} < 1.05$ (Muthén & Asparouhov, 2012). We report posterior medians with equal-tailed 95% CIs. Model fit was summarized by the posterior predictive $p$-value (PPP) and the 95% posterior predictive check credible interval (PPC CI) for the observed–replicated discrepancy, as well as the Deviance Information Criterion (DIC) and the effective number of parameters (pD). All models converged ($\hat{R} < 1.05$) and posterior predictive checks indicated adequate fit (PPP values near .50).

---

[2] Residual degrees of freedom were defined as $N_{L1} - J - p$, where $N_{L1}$ is the number of L1 units, $J$ is the number of clusters, and $p$ is the number of fixed L1 predictors.

**Real Data Results**

***Influence of Informative Priors on Variance Estimates***

Informative *IG* priors significantly influenced variance component estimation in both the full sample and subsample, especially in small-sample conditions. This aligns with the expectation that informative priors stabilize variance estimates in small samples. Main results are presented here, with extended tables and figures available in Appendix S3.

**Level-2 Variance Estimates.** Figure 3 (Panels A–B) shows the L2 intercept variance (i.e., between-school variance in science scores) across different prior conditions. In the full sample, point estimates clustered around 152–155, regardless of prior informativeness. However, the CIs were slightly narrower under the mode-matched informative $IG(a, b)$ priors (BI VarIGab, CI width = 44.09) compared to flat priors (BUI, CI width = 85.85), with the two combination models bracketing these values (Appendix S3, Table S6). In contrast, prior choice mattered substantially in the subsample. Flat and weak variance priors produced very wide or unstable posteriors (e.g., BUI, CI width = 360.13; $IG(.01,.01)$ yielded near-zero L2 variance with the lower bound at the truncation point). In comparison, the mode-matched $IG(a, b)$ prior stabilized the estimate near the full-sample target (means = 152.74) and reduced interval width by more than half (width = 139–140; see Table 4).

**Level-1 Residual Variance.** In the full sample, median posterior estimates were consistent across all prior conditions (approximate 2,134), with modest interval tightening under $IG(a, b)$ priors (CI width = 101.88) relative to flat priors (CI width = 125.73) (see Appendix S3, Table S6 and Figure S5 – Panel A). In the subsample, the $IG(a, b)$ prior again yielded more stable and narrower CIs (width = 820.44–822.65) compared to flat priors (CI width = 995.84), while the median posterior estimates remained close to 2060–2066 across conditions (see Table 4 and Appendix S3, Figure S6 – Panel A). These results demonstrate that informative, mode-matched $IG(a, b)$ prior significantly stabilized the L2 variance in the small-sample context and modestly tightened the L1 residual variance in both samples. Weak variance priors (e.g., $IG(.01,.01)$) risked pathological underestimation of the L2 variance in the subsample.

### *Influence of Informative Priors on Fixed Effects*

In contrast to their more substantial impact on variance components, the effect of informative $IG$ priors on fixed effect estimates was relatively modest. However, the impact was non-negligible in the subsample, particularly in terms of posterior precision.

**Full Sample.** Across all conditions (flat priors, weak priors, and informative priors), posterior estimates for the intercept (523.19−523.24; CI width = 3.62–3.65), within-school math score (L1 predictor = 59.12; CI width = 1.90), between-school math score (L2 predictor = 64.21−64.22; CI width = 3.75–3.77), and the cross-level interaction (= −2.72; CI width = 2.21) remained stable and consistent in both magnitude and direction. CIs overlapped substantially across conditions, and posterior median estimates varied only slightly. These results are expected, as the large sample size provides sufficient information for robust estimation of fixed effects, regardless of prior strength. See Appendix S3, Figure S5 (Panels B–E) and Table S5.

**Subsample.** Point estimates for fixed effects were stable across priors, but interval widths showed greater sensitive. For the intercept and L2 math effect, the weak variance prior $IG(.01,.01)$ (with or without weakly informative slope priors) yielded the narrowest intervals (e.g., intercept CI width = 13.13–13.15; L2 math CI widths = 11.91–11.95). In contrast, the mode-matched $IG(a, b)$ prior produced similar centers but wider intervals for these slopes (e.g., intercept CI width = 18.16–18.24; L2 math CI width = 18.04–18.05). The L1 math effect remained precise across all prior conditions (CI width = 14.34–14.65). The interaction was imprecise in all conditions (CI width = 15.47–15.79), with its median near −8 showing minimal variation across prior specifications (Appendix S3, Figure S6 Panels B–E; Table 4).

The key takeaway is that fixed-effect medians were robust to prior choice. In the small sample, slope interval widths varied modestly by variance prior type: $IG(.01,.01)$ yielded the tightest slope CIs, while mode-matched $IG(a, b)$ emphasized stabilizing variance components rather than constraining the slopes themselves.

### Comparison between Full sample and Subsample Results

The comparison between the full sample and subsample results highlights that prior choice had minimal effect on fixed effects but pronounced consequences for the L2 variance component see again Figure 3, Panels A–B). In the full sample, variability in the L2 parameter estimates was consistent irrespective of priors used, with only modest gains in precision under mode-matched $IG(a, b)$ priors (e.g., CI width = 44.09 for $IG(a, b)$ vs. 85.85 for flat priors). This consistency is expected given the large sample size of schools, where data typically outweighs the influence of the prior. In the subsample, however, prior sensitivity was much more pronounced. Flat and weak variance priors produced highly unstable or diffuse L2 variance estimates (e.g., BUI width = 360; $IG(.01, .01)$ centered near zero), whereas the mode-matched $IG(a, b)$ anchored the posterior near the full-sample reference value. As shown in Figure 3 – Panel B, the red dotted line marks the full-sample posterior median under the flat prior (BUI). The subsample $IG(a, b)$ median fell slightly below this benchmark, but their CIs were narrower (CI width 139–140) and closely centered around it, indicating increased precision compared to uninformative priors. Unlike in the simulation study, where the true population parameter was known, the red line here represents a sample-based benchmark from the full sample. Its role is to illustrate that informative priors in small-$J$ settings can pull estimates toward empirically grounded values while improving precision.

In sum, priors had negligible impact on fixed effects in the full sample but were consequential in the subsample, where informative variance priors produced stable L2 variance estimates. Fixed effects remained robust across all priors. This highlights the importance of using informative variance priors when working with multilevel data with relatively few higher-level units.

### Posterior Predictive Checks and Information Criteria

Posterior predictive checks indicated adequate fit for all models. In the full sample, PPP values ranged from 0.515 to 0.574, and the 95% PPC CIs for the observed $-$ replicated $\chi^2$ statistic consistently straddled zero across priors. DIC values differed by less than 2 units (91,543–91,545) and pD was stable at 186–187. In the subsample, PPP values were again close to 0.50 (0.490–0.534), PPC CIs included zero, and DIC differences were minimal (1,577–1,579). The $IG(.01,.01)$ models had slightly smaller pD (= 7.5) than the $IG(a, b)$ models (= 12.3), reflecting minor differences in effective complexity but not indicating any substantive fit advantage for a particular prior family. Extended results are available in Appendix S3, Table S6.

### Sensitivity Analysis

To assess the robustness of the results obtained under informative $IG$ variance priors, a sensitivity analysis was conducted by perturbing prior strength while holding the prior mode constant. Specifically, the shape parameter $a$ was varied by ±25% around the baseline values used in the main analysis (BI VarIGab and BI Coef + VarIGab). The corresponding scale parameter $b$ was recalculated each time to preserve the prior mode $b/(a + 1)$. This approach allowed us to examine the extent to which moderate misspecification of prior informativeness affected posterior inferences, without shifting the central tendency of the prior (see Appendix S2; Depaoli & van de Schoot, 2017; McNeish, 2016; van Erp et al., 2018).

Tables S7 and S8 in Appendix S3 provide the posterior summaries of fixed and random effects under the six prior-strength conditions. In the full sample, posterior median and CIs for all parameters were virtually unchanged, indicating that large sample size dominated the priors. In the subsample, small shifts in CI widths were observed, particularly for the L2 variance component, but the posterior median remained stable. The magnitude of these shifts was minor, suggesting the results were not unduly sensitive to moderate changes in prior informativeness.

Across all variations in prior strength, posterior predictive fit indices remained stable (Appendix S3, Table S9). PPP values remained close to 0.50, and the 95% PPC CIs consistently included zero, indicating no degradation in overall model fit. DIC and pD values showed only minor fluctuations, further suggesting limited sensitivity to prior strength. In the full sample (273 schools), posterior median and 95% CIs were essentially unchanged across baseline, weaker, and stronger prior conditions, reflecting the dominance of the data at large $J$. In the subsample (30 schools), CI widths shifted slightly, particularly for the L2 variance component, but posterior medians remained stable. For example, under $IG(a, b)$ with $a$ varied by ±25%, the L2 variance CI widened slightly (125–159) but remained centered on the same median (=153). Fixed effects were virtually unaffected, with medians and intervals nearly identical across conditions.

Taken together, these results demonstrate that the proposed mode-matched $IG$ priors are robust to moderate variation in prior strength. The sensitivity analysis confirms that scaling $a$ up or down by 25% has negligible influence on posterior conclusions. Anchoring priors to empirical benchmarks and calibrating their strength to sample size thus provides a flexible and stable strategy for multilevel applications, particularly in small-sample contexts where variance components are most vulnerable to prior influence.

## Discussion

The present study addresses the challenge of estimating variance components in Bayesian MLMs when the number of groups is small, by using informative *IG* priors. Overall, the findings provide strong evidence that carefully constructed informative variance priors can meaningfully improve estimation and inference in these contexts. In both simulations and the TIMSS example, we observed that informative variance priors substantially improved estimation relative to weakly informative or default choices. Specifically, they reduced bias, narrower CIs, and stabilized posterior estimates of random-effect variances. From a practical perspective, this means researchers working with sparse higher-level samples can obtain more credible and interpretable variance estimates by incorporating external knowledge into the prior.

The applied example with TIMSS data illustrated these benefits in practice. With only 30 schools, maximum likelihood estimates risked overstating or understating between-school variance, depending on sample idiosyncrasies. By anchoring the prior on information from the full dataset, the Bayesian analysis produced a stable estimate of modest between-school variability in science achievement, reflecting neither collapse to zero nor implausible inflation. Stabilizing the variance estimate also improved inference for regression coefficients and cross-level interactions, yielding tighter intervals and more definitive tests. This demonstrates how informative priors can help applied researchers draw more reliable substantive conclusions even in studies constrained by small numbers of higher-level units.

It is important to emphasize that the informative priors are not "cheating" or biasing results in an arbitrary direction; rather, but the result of incorporating plausible prior knowledge, consistent with Bayesian principles. When the prior reflects true or approximately true values, it meaningfully improves estimation efficiency compared to uninformative priors. At the same time, our analyses highlight that Bayesian inference is only as trustworthy as the assumptions embedded in the prior. If the prior is misspecified, estimates may be pulled away from the truth. For this reason, careful prior elicitation and validation are essential. Our sensitivity analysis, which varied prior strength by ±25%, suggested robustness to moderate deviations in informativeness. In applied work, researchers should perform similar checks by varying prior parameters (e.g., strength or mode) within a reasonable range, ensuring that conclusions do not hinge on fragile prior choices.

**Practical implications and recommendations**

The current study has several practical takeaways for applied researchers:

Firstly, the use of informative priors can be highly beneficial when the number of groups is small. Rather than relying on ML (which can underperform) or default vague priors (which can yield overly uncertain results), researchers should incorporate external information about group-level variance. Even moderate informativeness can provide stabilization. For instance, setting an $IG$ prior with shape $a$ 2–5 and mode equal to a reasonable guess will introduce some gentle regularization that prevents extreme estimates without strongly dictating the result.

Priors should be anchored in either empirical evidence or well-founded expectations. A key strength of the proposed approach is its transparency: priors are parameterized in terms of a mode (the most plausible variance value) and an effective sample size (reflected through the shape parameter, $a$). We encourage researchers to think explicitly in these terms: *"What is a sensible variance value, and how confident am I in that judgment?"* The answer can then be translated directly into an $IG(a, b)$. For example, if prior research suggests that the between-group standard deviation is likely near 10, but values from 5 to 20 are plausible, this reasoning could be encoded as an *IG* prior with mode $10^2 = 100$ and a relatively small shape parameter ($a = 3$ or 4), which produces a wide 90% prior interval (approximately [25, 400]). Conversely, if a large, high-quality study has provided a reliable variance estimate, a stronger prior (larger $a$) would be appropriate to reflect greater certainty.

Researchers should avoid extremely uninformative priors for variance components, particularly in small samples. Flat (truly uniform, e.g., $p(\sigma^2) \propto 1$) or diffuse priors (technically proper but extremely wide, e.g., $IG(.001, .001)$ for variance or $N(0, 10,000)$ for regression coefficients) can lead to non-convergence and distorted posteriors. In our simulations, such priors occasionally produced aberrantly large variance estimates that inflated RMSE and reduced stability (Gelman, 2006; McNeish, 2016a). A safer strategy is to use weakly informative priors that constrain extreme values without overwhelming the likelihood. Examples include a half-*t* prior with a reasonable scale or an *IG* with $a > 2$ and $b$ calibrated by mode or a high quantile (e.g., ensuring 90% prior mass below a plausible upper bound). Such priors reduce the risk of implausibly small or large variance estimates while still allowing the data to guide inference.

Flat priors can lead to non-convergence or distorted posteriors by allowing the model to explore implausible parameter regions (Gelman, 2006; McNeish, 2016a). Our simulations suggested such priors occasionally produced aberrantly large variance estimates that inflate RMSE and reduce stability. A safer strategy is to use weakly informative priors that constrain extreme values without overwhelming the likelihood. Examples include a half-*t* prior with a reasonable scale or an *IG* with $a > 2$ and $b$ calibrated by mode or a high quantile (e.g., ensuring 90% prior mass below a plausible upper bound). Such priors reduce the risk of implausibly small or large variance estimates while still allowing the data to drive inference. In short, unless there is strong justification otherwise, researchers should replace diffuse variance priors with weakly informative alternatives to improve stability and interpretability.

In applied work, this means that researchers should not purely rely on software defaults, which often specify diffuse or improper priors that are ill-suited for small-sample MLMs. Instead, defaults should be checked and, where necessary, replaced with priors grounded in empirical evidence or substantive expertise. The approach presented in this paper offers a systematic way to do so for variance components: by deriving informative *IG* priors from scaled chi-square distributions and matching the prior mode to a plausible variance value, researchers can specify priors in terms of interpretable quantities (a variance mode and a shape parameter reflecting certainty).

The scale of the outcome and predictors is important when specifying variance priors. As we observed in our analyses, priors such as *IG*(.01, .01) that might appear "vague" on one measurement scale can become unintentionally informative on another. For example, an outcome measured 0–100 with typical variances in the thousands requires a very different prior than an outcome scaled 0–1. In practice, researchers should either standardize variables or explicitly calibrate priors to the metric of the data, ensuring that the prior reflects knowledge on the same scale as the observed outcome.

Prior sensitivity analysis should be a routine component of Bayesian reporting. In this study, we evaluated sensitivity by varying the *IG* shape parameter (*a*) by ±25% while holding the prior mode fixed; posterior inferences were substantively unchanged across these settings. As an additional check, researchers can compare alternative prior families calibrated to similar central tendencies (e.g., replacing *IG* with a half-*t* on the standard deviation using a scale chosen to yield a comparable prior mode/mean) and assess whether posterior conclusions materially differ. If findings remain stable across reasonable prior specifications, this strengthens confidence in their robustness; if not, conclusions should be presented with appropriate caution and with explicit acknowledgment of their dependence on prior assumptions.

**Limitations and Future Research**

A key limitation of this study is that the simulation was conducted under a "best-case" scenario, with informative $IG(a, b)$ priors derived directly from the true population variances. While this design provides a clean benchmark, it does not reflect the uncertainty faced in applied settings where population values are unknown. Thus, the results should be interpreted as showing the potential of correctly specified informative priors, rather than evidence that they will always outperform weakly informative alternatives. To partially address this, the real-data analysis included a sensitivity check varying the $IG$ shape parameter by ±25% while fixing the mode. The results indicated that posterior inferences and fit indices were robust to such moderate deviations. Nevertheless, future research should explicitly model prior misspecification, for example, by centering priors away from the true variance or altering their spread to better understand how informative priors behave under realistic conditions

A related limitation is the empirical Bayes aspect of our real-data demonstration. We used the full sample's estimates to inform the small sample's prior. This practice is somewhat circular, as it uses the same data twice, and therefore is not a purely Bayesian approach. We did it here purely to illustrate the maximum-benefit scenario. In applied work, priors should ideally be informed by external data sources or domain expertise, and when empirical Bayes is used, the dependence between prior and data should be acknowledged and, where possible, adjusted for (Efron, 2012). Accordingly, our TIMSS results should be viewed as a demonstration rather than an endorsement of using full-sample estimates as priors for subsamples in a real analysis.

Another limitation is that we focused on a relatively simple random intercept model. Many applied multilevel analyses involve random slopes and covariance structures between random effects. Extending informative priors to covariance matrices is more complex. One approach is to apply mode-matched priors to each variance and an LKJ prior for correlations (Lewandowski et al., 2009). However, we did not address these scenarios, but currently limited to scalar variance parameters. Small-sample issues are often exacerbated in random slope models, particularly in estimating covariances (Browne & Draper, 2006; McNeish & Stapleton, 2016b). Future work should investigate informative priors for covariance components, potentially drawing on prior correlation estimates or cross-validation strategies.

A further limitation is that we restricted our evaluation to the *IG* family, leaving other widely discussed priors (e.g., half-*t*, half-Cauchy, PC; Gelman, 2006; Simpson et al., 2017) untested. Future simulation work should compare these approaches directly, calibrating them to have equivalent information content (e.g., same prior mean or CI) to determine how their differing tail behavior affects bias, efficiency, and robustness in MLMs.

Finally, our simulations were limited to balanced cluster sizes and normally distributed outcomes. In practice, researchers often face unbalanced designs or non-normal outcomes (e.g., binary responses). The role of informative priors under these conditions is less clear, though we expect the general pattern to hold: small numbers of groups remain problematic, and informative priors can help, but performance may differ. For instance, in logistic random-intercept models, very small cluster counts can lead to non-identification of variance components and separation issues, making prior information especially important (Gelman et al., 2008). Extending this approach to binary or categorical MLMs is therefore an important direction for future work.

In conclusion, this work demonstrates that incorporating well-informed priors for variance components is theoretically appealing and practically helpful for MLMs with limited higher-level sample sizes. By transforming substantive or empirical knowledge into a statistical prior thorough the chi-square → gamma → inverse-gamma framework, researchers can achieve more reliable estimates than otherwise possible. Our hope is that this approach and the provided guidelines encourage wider adoption of principled informative priors in MLMs, ultimately improving the robustness of conclusions drawn in fields like education, psychology, and beyond where hierarchical data structures and small samples are common.

**Acknowledgments**

The author would like to thank Dr. Elizabeth Sanders for valuable feedback and guidance during the development of this work.

# Appendix A

## Quick Identities and Practical Recipe for Mode-Matched Inverse-Gamma Priors

**Contents**
A1. Quick-reference formulas
A2. Practical recipe for constructing mode-matched *IG* priors
Note on full details and examples

### A1. Quick-reference formulas
• Chi-square-Gamma identify
$$\boldsymbol{\chi_\nu^2 \equiv \Gamma(\nu/2, 2)}$$
If $V = c \cdot \chi_\nu^2$ with $c > 0$, then $V \sim \Gamma(\nu/2, 2c)$.

• Reciprocal link between $\Gamma$ and IG
If $P \sim \Gamma(\alpha, \theta)$ then $X = 1/P$, then $X \sim IG(\alpha, \beta = 1/\theta)$

• Inverse-gamma (IG) facts (shape $a$, scale $b$)
pdf:
$$g(x \mid a, b) = \frac{b^a}{\Gamma(a)} x^{-(a+1)} \exp\left(-\frac{b}{x}\right), x > 0$$
mode: $b/(a + 1)$ for $\alpha > 1$
mean: $b/(a - 1)$ for $\alpha > 1$
variance: $b^2/[(a - 1)^2(a - 2)]$ for $a > 2$

### A2. Practical recipe for constructing mode-matched IG priors
1) Choose a target mode $m$ from prior research, pilot data, or a large related dataset.
2) Choose a prior strength via $\nu$ (effective degrees of freedom) and set $a = \nu/2$. Values between 4 and 40 (i.e., $a$ between 2 and 20) are often reasonable for weak-to-moderate informativeness.
3) Set $b = m(a + 1)$ so that mode $(IG) = b/(a + 1) = m$. Ensure $a > 1$ for a finite mean and $a > 2$ for a finite variance.
4) Check prior implications: compute the mean $b/(a - 1)$ and variance $b^2/[(a - 1)^2(a - 2)]$ (for $a > 2$), and optionally compute the 95% prior interval numerically or via prior predictive simulation.
5) Conduct sensitivity analysis by multiplying $a$ by 0.75 and 1.25 and resetting $b = m(a + 1)$ to keep the mode fixed at $m$. Report any substantive changes.

### Note on full details and examples
Full distributional definitions, derivations, and worked parameter examples are provided in the online Supplementary Materials, Appendix S1 (Mathematical Details & Worked Examples). Sensitivity specifications and Mplus code are provided in Appendix S2. Extended tables and figures appear in Appendix S3.

**Table 1**

*False Positive Rates for the Level-2 Intercept Coefficient and Raw Bias for the Level-2 Intercept and Variance by Prior Condition and Design Factors*

| Condition | False Positive Rate | | | | | | | L2 Intercept Coeff Raw Bias | | | | | | | L2 Intercept Variance Raw Bias | | | | | | |
|---|---|---|---|---|---|---|---|---|---|---|---|---|---|---|---|---|---|---|---|---|---|
| | ML | BUI | BI-N | BI-G01 | BI-Gab | BI-N+G01 | BI-N+Gab | ML | BUI | BI-N | BI-G01 | BI-Gab | BI-N+G01 | BI-N+Gab | ML | BUI | BI-N | BI-G01 | BI-Gab | BI-N+G01 | BI-N+Gab |
| **ICC = .01** | | | | | | | | | | | | | | | | | | | | | |
| *J* = 10 | | | | | | | | | | | | | | | | | | | | | |
| *M* = 5 | .00 | .00 | .03 | .03 | .03 | .03 | .03 | .01 | .01 | .04 | .04 | .02 | .04 | .02 | .00 | .13 | .03 | .03 | .01 | .03 | .01 |
| *M* = 30 | .01 | .01 | .02 | .02 | .03 | .02 | .03 | .00 | .00 | .02 | .02 | .02 | .02 | .02 | .00 | .02 | .01 | .01 | .00 | .01 | .00 |
| *J* = 30 | | | | | | | | | | | | | | | | | | | | | |
| *M* = 5 | .00 | .01 | .02 | .02 | .04 | .02 | .04 | .00 | -.02 | -.02 | -.02 | -.02 | -.02 | -.02 | .00 | .04 | .02 | .02 | .01 | .02 | .01 |
| *M* = 30 | .02 | .03 | .04 | .04 | .04 | .04 | .04 | .00 | .00 | -.01 | -.01 | -.01 | -.01 | -.01 | .00 | .01 | .01 | .01 | .00 | .01 | .00 |
| *J* = 100 | | | | | | | | | | | | | | | | | | | | | |
| *M* = 5 | .02 | .04 | .04 | .04 | .04 | .04 | .04 | .00 | .00 | .00 | .00 | .00 | .00 | .00 | .01 | .02 | .01 | .01 | .01 | .01 | .01 |
| *M* = 30 | .05 | .04 | .04 | .04 | .04 | .04 | .04 | .00 | .00 | .00 | .00 | .00 | .00 | .00 | .00 | .00 | .00 | .00 | .00 | .00 | .00 |
| **ICC = .20** | | | | | | | | | | | | | | | | | | | | | |
| *J* = 10 | | | | | | | | | | | | | | | | | | | | | |
| *M* = 5 | .04 | .00 | .05 | .05 | .01 | .05 | .01 | .04 | .02 | .03 | .03 | .07 | .03 | .07 | -.13 | .21 | -.06 | -.06 | .07 | -.06 | .07 |
| *M* = 30 | .13 | .01 | .06 | .06 | .03 | .06 | .03 | -.01 | .01 | .03 | .03 | .03 | .03 | .03 | -.08 | .16 | .01 | .01 | .06 | .01 | .06 |
| *J* = 30 | | | | | | | | | | | | | | | | | | | | | |
| *M* = 5 | .06 | .04 | .05 | .05 | .04 | .05 | .04 | -.01 | -.02 | -.02 | -.02 | -.02 | -.02 | -.02 | -.07 | .01 | -.05 | -.05 | .02 | -.05 | .02 |
| *M* = 30 | .08 | .03 | .04 | .04 | .03 | .04 | .03 | .00 | .02 | .01 | .01 | .02 | .01 | .02 | -.03 | .02 | .00 | .00 | .01 | .00 | .01 |
| *J* = 100 | | | | | | | | | | | | | | | | | | | | | |
| *M* = 5 | .07 | .05 | .05 | .05 | .05 | .05 | .05 | .00 | .00 | .00 | .00 | .00 | .00 | .00 | -.02 | .00 | -.01 | -.01 | .01 | -.01 | .01 |
| *M* = 30 | .06 | .07 | .07 | .07 | .07 | .07 | .07 | .00 | .00 | .00 | .00 | .00 | .00 | .00 | -.01 | .00 | -.01 | -.01 | .00 | -.01 | .00 |
| **ICC = .40** | | | | | | | | | | | | | | | | | | | | | |
| *J* = 10 | | | | | | | | | | | | | | | | | | | | | |
| *M* = 5 | .10 | .01 | .06 | .06 | .02 | .06 | .02 | .05 | .02 | .03 | .03 | .07 | .03 | .07 | -.28 | .41 | -.10 | -.10 | .14 | -.10 | .14 |
| *M* = 30 | .14 | .01 | .06 | .06 | .03 | .06 | .03 | -.01 | .02 | .03 | .03 | .04 | .03 | .04 | -.17 | .40 | .05 | .05 | .16 | .05 | .16 |
| *J* = 30 | | | | | | | | | | | | | | | | | | | | | |
| *M* = 5 | .07 | .05 | .06 | .06 | .05 | .06 | .05 | -.02 | -.02 | -.02 | -.02 | -.03 | -.02 | -.03 | -.10 | .03 | -.05 | -.05 | .04 | -.05 | .04 |
| *M* = 30 | .06 | .03 | .04 | .04 | .03 | .04 | .03 | .00 | .02 | .02 | .02 | .02 | .02 | .02 | -.06 | .04 | .00 | .00 | .04 | .00 | .04 |
| *J* = 100 | | | | | | | | | | | | | | | | | | | | | |
| *M* = 5 | .08 | .07 | .07 | .07 | .07 | .07 | .07 | .00 | .00 | .00 | .00 | .00 | .00 | .00 | -.02 | .00 | -.02 | -.02 | .00 | -.02 | .00 |
| *M* = 30 | .06 | .07 | .08 | .08 | .07 | .08 | .07 | .00 | .00 | .00 | .00 | .00 | .00 | .00 | -.02 | .00 | -.01 | -.01 | .00 | -.01 | .00 |
| *J* = 10 | **.07** | **.01** | **.05** | **.05** | **.03** | **.05** | **.03** | **.01** | **.01** | **.03** | **.03** | **.04** | **.03** | **.04** | **-.11** | **.22** | **-.01** | **-.01** | **.07** | **-.01** | **.07** |
| *J* = 100 | **.05** | **.06** | **.06** | **.06** | **.06** | **.06** | **.06** | **.00** | **.00** | **.00** | **.00** | **.00** | **.00** | **.00** | **-.01** | **.00** | **-.01** | **-.01** | **.00** | **-.01** | **.00** |
| **Total** | **.06** | **.03** | **.05** | **.05** | **.04** | **.05** | **.04** | **.00** | **.00** | **.01** | **.01** | **.01** | **.01** | **.01** | **-.05** | **.08** | **-.01** | **-.01** | **.03** | **-.01** | **.03** |

*Note.* Cell means above based on *N* = 1,800 replications (200 replications per condition x nine crossed levels of L1 and L2 Total *R*-squared values). Freq = Frequentist, BUI = Bayesian uninformative (flat) priors, BI = Bayesian with informative priors; N = Normal Distribution used for predictor coefficient priors, VarIG01 = IG(.01, .01) used for variance component priors, VARIGab = IG(*a*,*b*) used for variance component priors, with *a* and *b* set to have a mode equal to the true variance component value.

**Table 2**

*False Positive Rate and Correct Positive for L2 Predictor Slope Coefficient, Collapsed across L1 Predictor Effects*

| Condition | False Positive Rate | | | | | | | Correct Positive (L2 Coeff Rsq = .10) | | | | | | | Correct Positive (L2 Coeff Rsq = .25) | | | | | | |
|---|---|---|---|---|---|---|---|---|---|---|---|---|---|---|---|---|---|---|---|---|---|
| | ML | BUI | BI-N | BI-G01 | BI-Gab | BI-N+G01 | BI-N+Gab | ML | BUI | BI-N | BI-G01 | BI-Gab | BI-N+G01 | BI-N+Gab | ML | BUI | BI-N | BI-G01 | BI-Gab | BI-N+G01 | BI-N+Gab |
| **ICC = .01** | | | | | | | | | | | | | | | | | | | | | |
| *J* = 10 | | | | | | | | | | | | | | | | | | | | | |
| *M* = 5 | .00 | .01 | .07 | .07 | .06 | .07 | .06 | .00 | .02 | .08 | .08 | .07 | .08 | .07 | .01 | .02 | .09 | .09 | .07 | .09 | .07 |
| *M* = 30 | .02 | .01 | .04 | .04 | .05 | .04 | .05 | .03 | .03 | .07 | .07 | .10 | .07 | .10 | .03 | .05 | .09 | .09 | .18 | .09 | .18 |
| *J* = 30 | | | | | | | | | | | | | | | | | | | | | |
| *M* = 5 | .01 | .09 | .10 | .10 | .10 | .10 | .10 | .02 | .11 | .12 | .12 | .12 | .12 | .12 | .02 | .12 | .13 | .13 | .15 | .13 | .15 |
| *M* = 30 | .04 | .06 | .07 | .07 | .08 | .07 | .08 | .07 | .17 | .19 | .19 | .20 | .19 | .20 | .09 | .27 | .28 | .28 | .33 | .28 | .33 |
| *J* = 100 | | | | | | | | | | | | | | | | | | | | | |
| *M* = 5 | .03 | .16 | .17 | .17 | .16 | .17 | .16 | .03 | .21 | .21 | .21 | .21 | .21 | .21 | .04 | .23 | .23 | .23 | .25 | .23 | .25 |
| *M* = 30 | .03 | .09 | .08 | .08 | .08 | .08 | .08 | .12 | .35 | .33 | .33 | .33 | .33 | .33 | .21 | .60 | .58 | .58 | .61 | .58 | .61 |
| **ICC = .20** | | | | | | | | | | | | | | | | | | | | | |
| *J* = 10 | | | | | | | | | | | | | | | | | | | | | |
| *M* = 5 | .04 | .01 | .06 | .06 | .02 | .06 | .02 | .07 | .05 | .18 | .18 | .07 | .18 | .07 | .12 | .14 | .34 | .34 | .24 | .34 | .24 |
| *M* = 30 | .16 | .01 | .07 | .07 | .01 | .07 | .01 | .27 | .06 | .17 | .17 | .09 | .17 | .09 | .51 | .29 | .53 | .53 | .42 | .53 | .42 |
| *J* = 30 | | | | | | | | | | | | | | | | | | | | | |
| *M* = 5 | .09 | .08 | .12 | .12 | .08 | .12 | .08 | .18 | .35 | .38 | .38 | .33 | .38 | .33 | .37 | .68 | .69 | .69 | .68 | .69 | .68 |
| *M* = 30 | .07 | .06 | .08 | .08 | .06 | .08 | .06 | .43 | .64 | .67 | .67 | .65 | .67 | .65 | .88 | .97 | .98 | .98 | .98 | .98 | .98 |
| *J* = 100 | | | | | | | | | | | | | | | | | | | | | |
| *M* = 5 | .10 | .06 | .07 | .07 | .06 | .07 | .06 | .44 | .67 | .67 | .67 | .67 | .67 | .67 | .84 | .99 | .99 | .99 | .99 | .99 | .99 |
| *M* = 30 | .04 | .06 | .07 | .07 | .06 | .07 | .06 | .90 | .99 | .99 | .99 | .99 | .99 | .99 | 1.00 | 1.00 | 1.00 | 1.00 | 1.00 | 1.00 | 1.00 |
| **ICC = .40** | | | | | | | | | | | | | | | | | | | | | |
| *J* = 10 | | | | | | | | | | | | | | | | | | | | | |
| *M* = 5 | .10 | .03 | .08 | .08 | .02 | .08 | .02 | .18 | .06 | .22 | .22 | .07 | .22 | .07 | .31 | .24 | .49 | .49 | .38 | .49 | .38 |
| *M* = 30 | .13 | .03 | .05 | .05 | .02 | .05 | .02 | .30 | .07 | .13 | .13 | .07 | .13 | .07 | .57 | .30 | .53 | .53 | .42 | .53 | .42 |
| *J* = 30 | | | | | | | | | | | | | | | | | | | | | |
| *M* = 5 | .10 | .06 | .08 | .08 | .06 | .08 | .06 | .31 | .55 | .57 | .57 | .53 | .57 | .53 | .67 | .91 | .93 | .93 | .91 | .93 | .91 |
| *M* = 30 | .07 | .07 | .09 | .09 | .07 | .09 | .07 | .49 | .73 | .76 | .76 | .75 | .76 | .75 | .92 | .99 | 1.00 | 1.00 | .99 | 1.00 | .99 |
| *J* = 100 | | | | | | | | | | | | | | | | | | | | | |
| *M* = 5 | .09 | .07 | .08 | .08 | .07 | .08 | .07 | .72 | .93 | .94 | .94 | .93 | .94 | .93 | .99 | 1.00 | 1.00 | 1.00 | 1.00 | 1.00 | 1.00 |
| *M* = 30 | .05 | .07 | .07 | .07 | .07 | .07 | .07 | .93 | 1.00 | 1.00 | 1.00 | 1.00 | 1.00 | 1.00 | 1.00 | 1.00 | 1.00 | 1.00 | 1.00 | 1.00 | 1.00 |
| ***J* = 10** | **.07** | **.02** | **.06** | **.06** | **.03** | **.06** | **.03** | **.14** | **.05** | **.14** | **.14** | **.08** | **.14** | **.08** | **.26** | **.17** | **.34** | **.34** | **.29** | **.34** | **.29** |
| ***J* = 100** | **.06** | **.08** | **.09** | **.09** | **.08** | **.09** | **.08** | **.52** | **.69** | **.69** | **.69** | **.69** | **.69** | **.69** | **.68** | **.80** | **.80** | **.80** | **.81** | **.80** | **.81** |
| **Total** | **.06** | **.06** | **.08** | **.08** | **.06** | **.08** | **.06** | **.31** | **.39** | **.43** | **.43** | **.40** | **.43** | **.40** | **.48** | **.54** | **.60** | **.60** | **.59** | **.60** | **.59** |

*Note.* Cell means above based on *N* = 600 replications (200 replications per condition x three levels of L1 Total *R*-squared values). ). Freq = Frequentist, BUI = Bayesian uninformative (flat) priors, BI = Bayesian with informative priors; N = Normal Distribution used for predictor coefficient priors, VarIG01 = IG(.01, .01) used for variance component priors, VARIGab = IG(*a*,*b*) used for variance component priors, with *a* and *b* set to have a mode equal to the true variance component value.

**Table 3**

*Steps for Deriving Informative Inverse-Gamma Priors for Variance Components (Full Sample, Bayesian Multilevel Analysis)*

| Step/Parameter | Quantity | Formula/Plug-in | L2 Intercept Variance | L1 Residual Variance |
|---|---|---|---|---|
| Inputs | Sample size | − | $J = 273$ schools | $N = 8698$ students |
| | df $k$ | L2: $k = J{-}1$ | 272 | − |
| | | L1: $k = N\text{-}J\text{-}p_w$ | − | 8423 |
| | Target SD | $s$ | 12.38 | 46.17 |
| | Target mode | $m = s^2$ | 153.27 | 2131.97 |
| Scaled Chi-square | Weight $W$ | $(s^2{+}2)/k$ | $(12.38^2{+}2)/272 = 0.571$ | $(46.17^2{+}2)/8423 = 0.253$ |
| | Mean | $Wk$ | 155.27 | 2133.97 |
| | Mode | $\max(Wk{-}2, 0)$ | 153.27 | 2131.97 |
| | Variance | $2Wk$ | 310.54 | 4267.94 |
| Gamma (rate = 1/2, scale = 2) | Shape $a$ | $Wk/2$ | 77.64 | 1066.99 |
| | Scale $\theta$ | Fixed = 2 | 2 | 2 |
| | Rate $\lambda$ | $1/\theta$ | 0.5 | 0.5 |
| | Mean | $a \cdot \theta$ | 155.27 | 2133.97 |
| | Mode | $(a{-}1) \cdot \theta\ (a > 1)$ | 153.27 | 2131.97 |
| | Variance | $a \cdot \theta^2$ | 310.54 | 4267.94 |
| | Mplus Prior (variance) | $G(a,b)$ | (77.64, 0.5) | (1066.99, 0.5) |
| Inverse-Gamma | Shape $a$ | (mean+mode)/2 | $(155.27{+}153.27)/2 = 154.27$ | $(2133.97{+}2131.97)/2 = 2132.97$ |
| | Scale $b$ | mean$(\alpha{-}1)$ | $155.27{\times}(154.27{-}1) = 23{,}798.54$ | $2133.97{\times}(2132.97{-}1) = 4{,}549{,}568.55$ |
| | Mean | $b/(a{-}1)$ | 155.27 | 2133.97 |
| | Mode | $b/(a{+}1)$ | 153.27 | 2131.97 |
| | Mplus Prior (variance) | $IG(a, b)$ | $IG(154.27, 23{,}798.54)$ | $IG(2132.97, 4{,}549{,}568.55)$ |

*Note.* This table illustrates the derivation of informative inverse-gamma (*IG*) priors for variance components in Bayesian multilevel models. The process begins with the chi-square distribution (theoretical sampling distribution for variances), applies a scaling factor *W* to align with observed variances, translates to the equivalent gamma distribution, and finally to the inverse-gamma. For Level-2 (between-school) variance we use $k = J - 1 = 272$. For Level-1 (residual) variance we use the residual degrees of freedom $k = N - J - p_w = 8698{-}273{-}2{=}8423$, where $p_w$ is the number of fixed Level-1 predictors. Changing $k$ only

affects the scaling weight $W$. $IG(a, b)$ uses scale $b$ (not rate). Parameters $a$ (shape) and $b$ (scale) were chosen to preserve the mode at the observed variance estimate from the full sample. These priors were specified in Mplus models as resid_between ~ IG(154, 23799) for the L2 intercept variance and resid_within ~ IG(2133, 4549569) for the L1 residual variance for full sample.

**Table 4**

*Parameter Estimates for 2-Level Bayesian Multilevel Modeling Using Inverse Gamma Prior Informativeness (Subsample)*

| | BUI | | | | | BI Coef N | | | | | BI VarIG01 | | | | |
|---|---|---|---|---|---|---|---|---|---|---|---|---|---|---|---|
| *Fixed Effects* | *Est* | *(SE)* | *LB* | *UB* | *Width* | *Est* | *(SE)* | *LB* | *UB* | *Width* | *Est* | *(SE)* | *LB* | *UB* | *Width* |
| Intercept (Mean) | 518.57 | (4.34) | 509.22 | 526.85 | 17.63 | 519.49 | (4.34) | 510.15 | 527.83 | 17.68 | 520.87 | (3.42) | 512.35 | 525.50 | 13.15 |
| L1 Math (*Z*) | 64.60 | (3.66) | 57.26 | 71.91 | 14.65 | 64.68 | (3.66) | 57.36 | 72.01 | 14.65 | 64.66 | (3.64) | 57.48 | 71.82 | 14.34 |
| L2 Math (*Z*) | 72.01 | (4.29) | 63.12 | 80.70 | 17.59 | 72.12 | (4.29) | 63.20 | 80.89 | 17.68 | 70.30 | (3.88) | 66.63 | 78.57 | 11.95 |
| L1 x L2 Interact | -7.95 | (4.11) | -15.88 | -0.09 | 15.79 | -7.95 | (4.11) | -15.86 | -0.09 | 15.76 | -7.95 | (4.07) | -15.65 | -0.18 | 15.47 |
| *Random Effects* | *Var* | *(SE)* | *LB* | *UB* | *Width* | *Var* | *(SE)* | *LB* | *UB* | *Width* | *Var* | *(SE)* | *LB* | *UB* | *Width* |
| Intercept (L2) | 79.25 | (103.76) | 5.55 | 365.68 | 360.13 | 78.85 | (103.55) | 5.69 | 364.96 | 359.27 | 0.43 | (19.67) | 0.01 | 65.78 | 65.76 |
| Residual (L1) | 2047.95 | (257.16) | 1600.10 | 2595.94 | 995.84 | 2047.17 | (257.25) | 1599.70 | 2599.81 | 1000.12 | 2060.24 | (248.46) | 1620.94 | 2582.28 | 961.34 |
| | BI VarIGab | | | | | BI Coef + VarIG01 | | | | | BI Coef + VarIGab | | | | |
| *Fixed Effects* | *Est* | *(SE)* | *LB* | *UB* | *Width* | *Est* | *(SE)* | *LB* | *UB* | *Width* | *Est* | *(SE)* | *LB* | *UB* | *Width* |
| Intercept (Mean) | 518.49 | (4.60) | 509.39 | 527.55 | 18.16 | 521.56 | (3.42) | 513.10 | 526.24 | 13.13 | 519.50 | (4.59) | 510.36 | 528.60 | 18.24 |
| L1 Math (*Z*) | 64.67 | (3.64) | 57.40 | 71.82 | 14.42 | 64.74 | (3.64) | 57.54 | 71.91 | 14.36 | 64.75 | (3.64) | 57.48 | 71.88 | 14.40 |
| L2 Math (*Z*) | 72.20 | (4.60) | 62.89 | 80.94 | 18.05 | 70.36 | (3.87) | 66.73 | 78.64 | 11.91 | 72.33 | (4.60) | 63.01 | 81.05 | 18.04 |
| L1 x L2 Interact | -7.96 | (4.07) | -15.57 | -0.08 | 15.49 | -7.95 | (4.07) | -15.66 | -0.16 | 15.49 | -7.95 | (4.07) | -15.56 | -0.08 | 15.48 |
| *Random Effects* | *Var* | *(SE)* | *LB* | *UB* | *Width* | *Var* | *(SE)* | *LB* | *UB* | *Width* | *Var* | *(SE)* | *LB* | *UB* | *Width* |
| Intercept (L2) | 152.74 | (36.14) | 100.51 | 239.94 | 139.43 | 0.44 | (19.98) | 0.01 | 66.04 | 66.03 | 152.78 | (36.15) | 100.71 | 240.76 | 140.05 |
| Residual (L1) | 2065.87 | (213.64) | 1682.07 | 2504.72 | 822.65 | 2059.77 | (248.79) | 1622.62 | 2588.78 | 966.17 | 2065.42 | (213.66) | 1682.09 | 2502.53 | 820.44 |

*Note.* Analysis involved $N = 150$ students (Level 1, L1) within 30 schools (Level 2, L2). Any acronyms for outcomes/predictors here. All lower-level (L1) predictors were cluster-mean centered, and continuous predictors were standardized. BUI = Bayesian uninformative (flat) priors, BI = Bayesian with informative priors; N = Normal Distribution used for predictor coefficient priors, VarIG01 = $IG(.01,.01)$ used for variance component priors, VARIGab = $IG(a,b)$ used for variance component priors; $a$ and $b$ are initially calculated from chi-square distributions, transformed to gamma, and then to inverse gamma.

**Figure 1**

*Comparison of $\chi^2$, Gamma, and Inverse-Gamma Priors Aligned to a Common Mode*

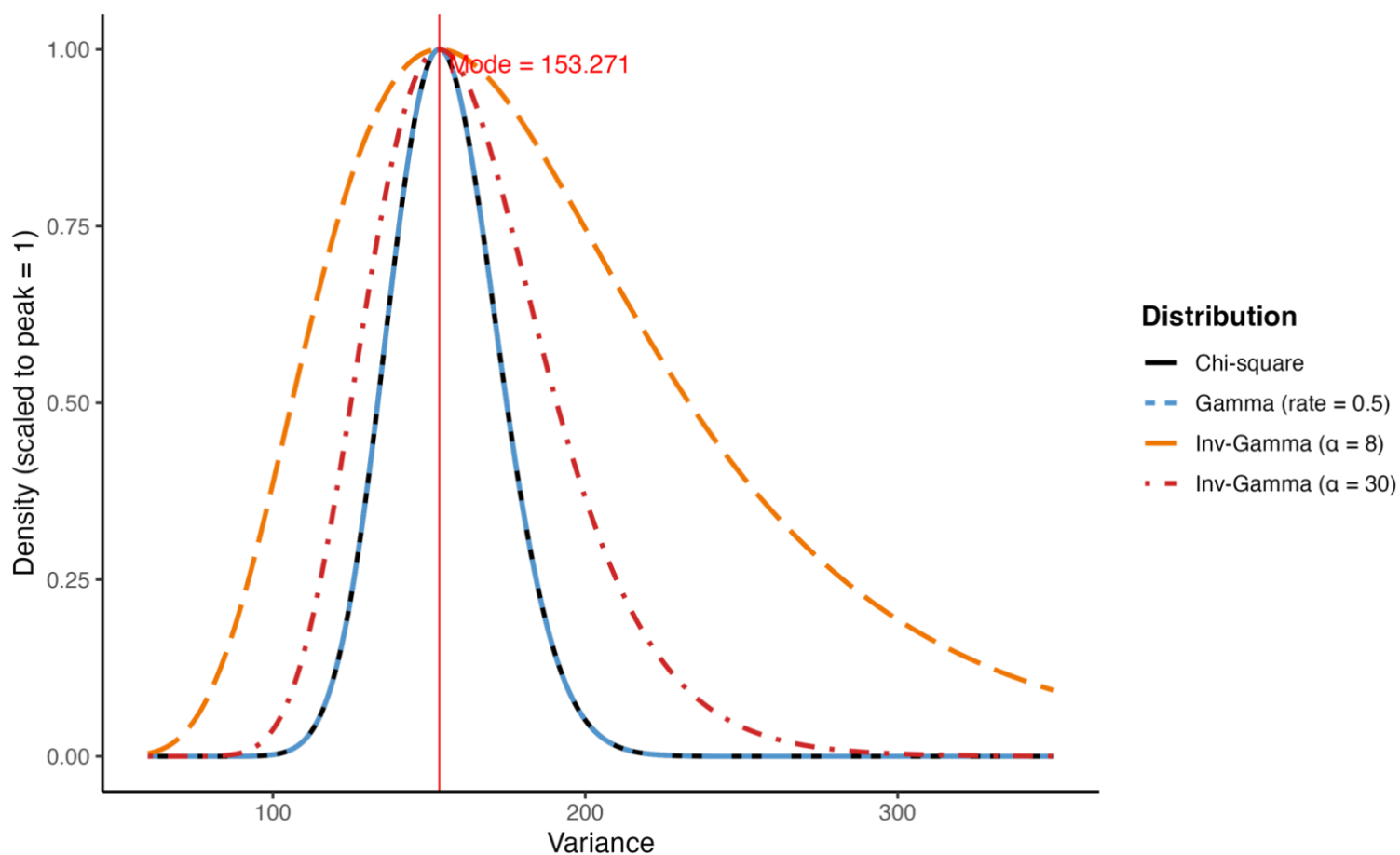

*Note.* Curves show three prior families parameterized on the variance scale to share the same prior mode at $m = 153.27$ (red vertical line). The inverse-gamma prior is $IG(a = 16.95, b = 2751.21)$, where the mode equals $b/(a + 1)$. The $\chi^2$ curve is a scaled chi-square with scale $c$ and degrees of freedom $\nu$ chosen so that $c(\nu - 2) = m$ (for $\nu \geq 2$). The gamma curve is parameterized on the variance scale with shape $k$ and scale $\theta$ chosen so that $(k - 1)\theta = m$. Densities are normalized for visual comparability; despite sharing the same mode, the families differ in tail behavior (the inverse-gamma exhibits the heaviest right tail). Abbreviations: $\chi^2$ = chi-square; $\Gamma$ = gamma; $IG$ = inverse-gamma.

**Figure 1**

*False Positive Rates for L2 Intercept Coefficient by Design Condition, Collapsed across L1 Predictor Effects*

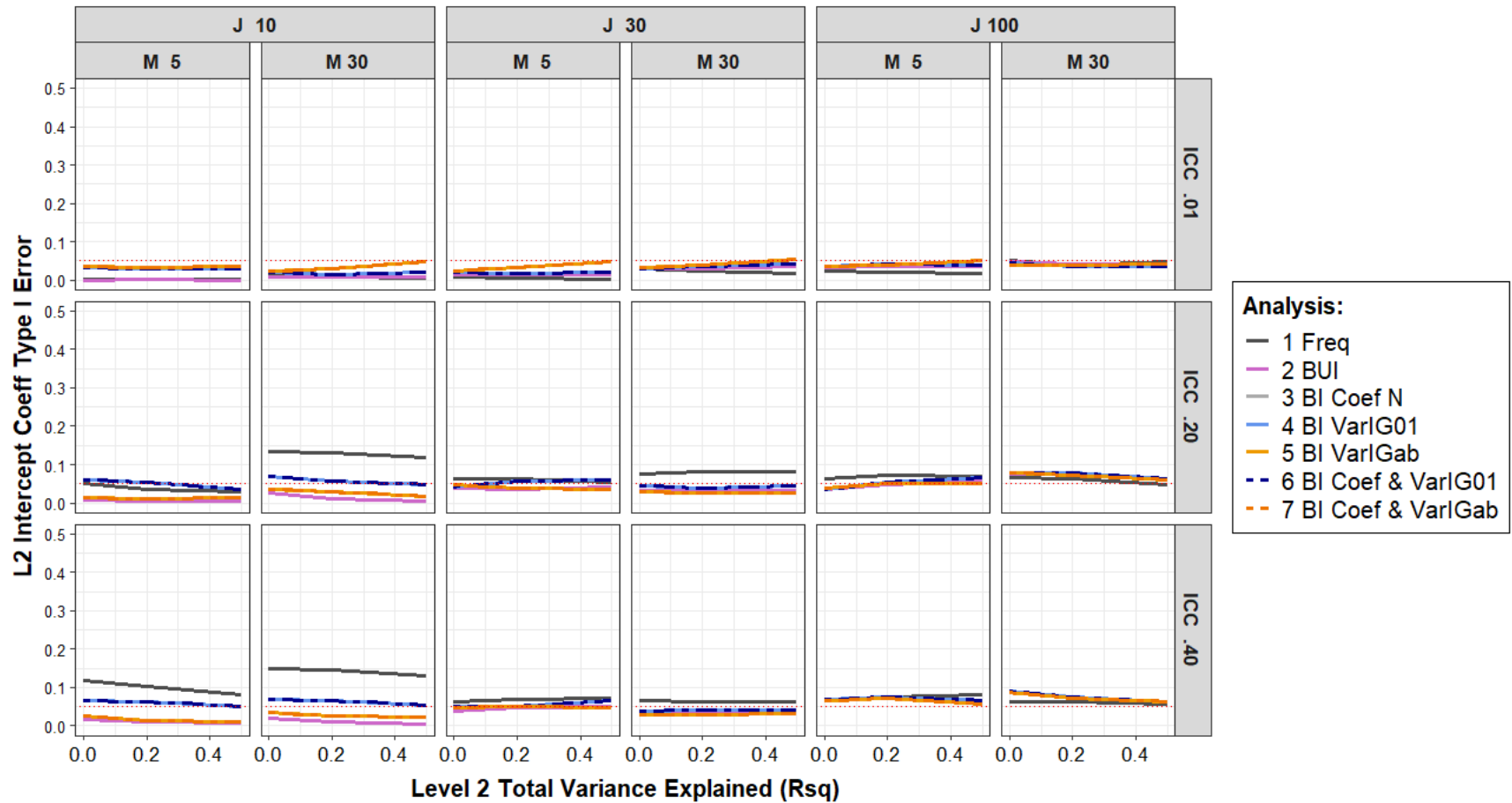

*Note.* False Positive rates are for all x-axis levels because intercept was set to zero (nominal false positive rate is .05). Freq = Frequentist, BUI = Bayesian uninformative (flat) priors, BI = Bayesian with informative priors; N = Normal Distribution used for predictor coefficient priors, VarIG01 = IG(.01,.01) used for variance component priors, VARIGab = IG($a$,$b$) used for variance component priors, with $a$ and $b$ set to have a mode equal to the true variance component value.

**Figure 2**

*Raw Bias Results for L2 Intercept Residual Variance by Design Condition, Collapsed across L1 Predictor Effects*

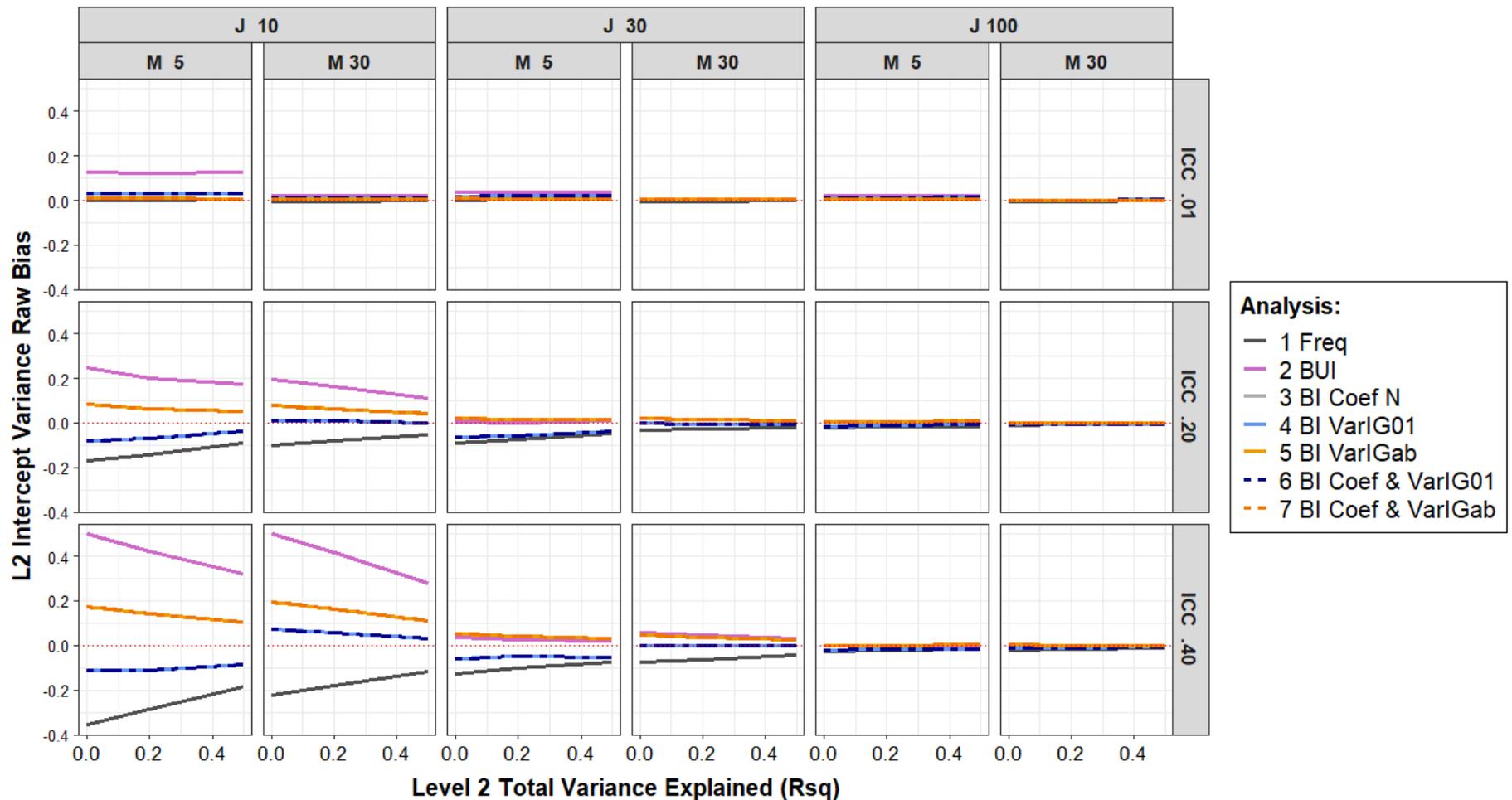

*Note.* Raw bias is difference between observed and population values. Freq = Frequentist, BUI = Bayesian uninformative (flat) priors, BI = Bayesian with informative priors; N = Normal Distribution used for predictor coefficient priors, VarIG01 = IG(.01,.01) used for variance component priors, VARIGab = IG($a$,$b$) used for variance component priors, with $a$ and $b$ set to have a mode equal to the true variance component value.

**Figure 3**

*Estimated L2 Intercept Variance Using Inverse Gamma Prior Informativeness (Full Sample)*

Panel A: Estimated L2 Intercept Variance Using IG Informativeness
(Full Sample)

Panel B: Estimated L2 Intercept Variance Using IG Informativeness
(Subsample)

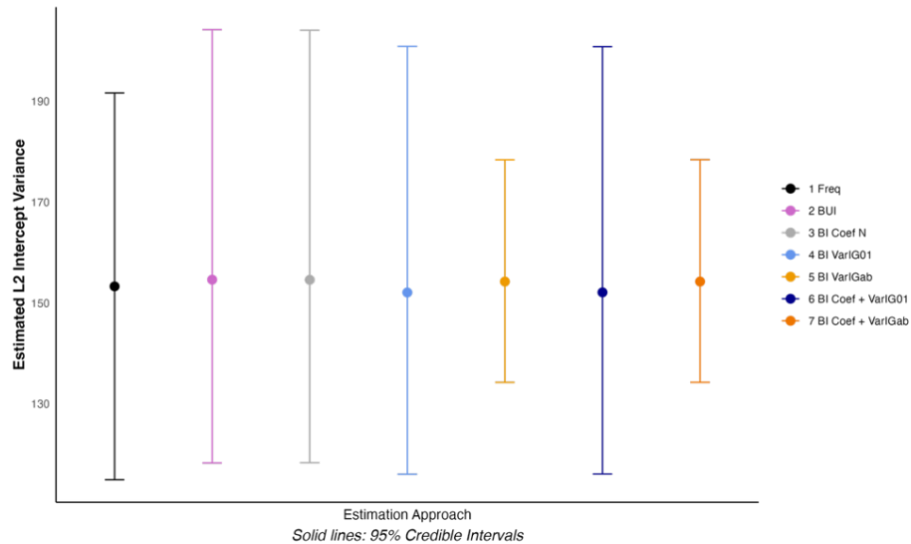
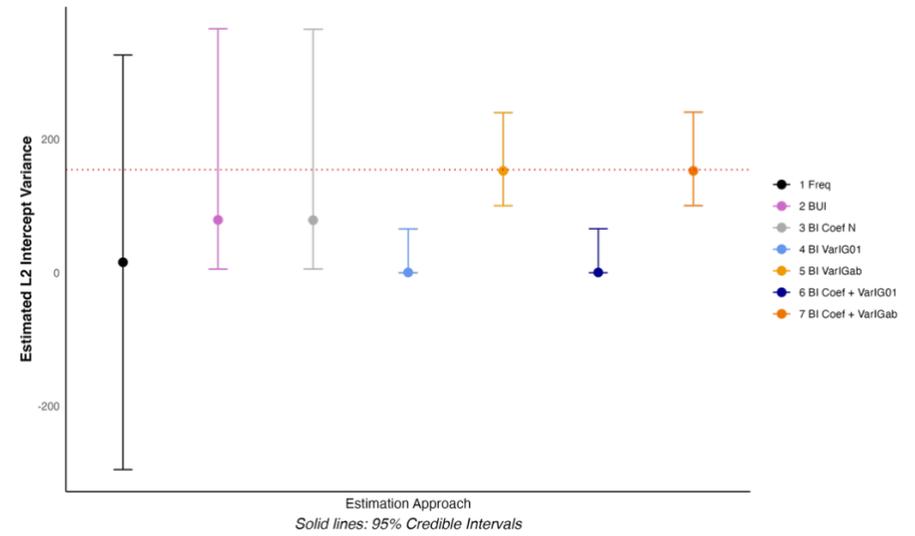

*Note.* Panel A shows the full-sample posterior medians and 95% credible intervals across priors; Panel B shows the subsample results. The red dotted line marks the full-sample posterior median from the Bayesian flat prior (BUI) model, used as a benchmark for subsample comparisons. BUI = Bayesian uninformative (flat) priors. BI = Bayesian with informative priors. N = Normal distribution used for predictor coefficient priors, VarIG01 = $IG(.01,.01)$ used for variance component priors, VARIGab = $IG(a,b)$ priors used for variance components; $a$ and $b$ are initially calculated from chi-square distributions, transformed to gamma, and then to inverse gamma.